\documentstyle{aipproc}

\begin{document}
\title{Blazars}

\author{Marek Sikora$^*$ and Greg Madejski$^{\dagger}$}
\address{$^*$N. Copernicus Astronomical Center\\
00716 Warsaw, Bartycka 18, Poland\\
$^{\dagger}$Stanford Linear Accelerator Center\\
Menlo Park, CA 94025, USA}

\maketitle

\begin{abstract}
In our review of the blazar phenomenon, we discuss blazar models, with
a focus on the following 
issues: sub-parsec jets and their environment; energy dissipation and particle 
acceleration; and radiative processes.  
\end{abstract}

\section*{Introduction}
Blazars are believed to be a sub-class of radio-loud AGNs with the 
relativistic jets 
pointing close to the line of sight, such that the entire 
electromagnetic spectra are 
dominated by non-thermal radiation produced in the jets. 
These spectra usually 
reveal two broad components, the low energy one  - peaking in the 
IR - to - X--ray range, and the high 
energy one - peaking in the MeV -- TeV range. Both spectral 
components are variable, particularly strongly in their high energy parts 
\cite{UMU,edel,weh,taka,taka2000}. 
Variability time scales range from years to a fraction 
of a day, where the rapid variability often takes the form of  high amplitude
flares. The most extreme events (with the largest amplitude 
and shortest time scales) 
have been recorded in $\gamma$--ray bands \cite{gai,mat}.  
Flares recorded in the GeV and optical bands in high luminosity
blazars \cite{wag,weh} and in the TeV and X-ray bands in low luminosity 
blazars \cite{mac,taka,samb,petry} 
appear to be correlated.  These are most likely produced by the 
high energy ends of the distributions of radiating particles, 
and this behavior suggests co-spatial production of high 
energy  and low energy components.  The flares are presumably 
produced  in shocks propagating down the 
jet with relativistic speeds and formed by colliding 
inhomogeneities \cite{rees,SBR,blaz,ghis}. 
However, present data do not allow us to exclude other 
possibilities, e.g., production of flares by inhomogeneities flowing
through and shocked in the reconfinement regions \cite{komis} or 
via collisions of jets with clouds \cite{chiang,laor}. 

In any case, the rapid, high amplitude variability events 
indicate that a significant
fraction of the blazar radiation is produced on sub-parsec scales and this
provides exceptional opportunity to explore jet properties in the vicinity
of the central engine. In order to advance such a study, one must 
first understand 
the dominant radiation processes operating in the sub-parsec jets. Until about 
10 years ago, the most viable radiation model for relativistic jets was the 
synchrotron-self-Compton (SSC) model \cite{konigl,margear,ghimar}.  
According to this model, the smooth, polarized and variable 
low energy component of blazar spectra is produced by the 
synchrotron mechanism, while the high energy component results from  
Comptonization of synchrotron radiation by the same population 
of electrons which produce the synchrotron component. However, this 
rather simple picture changed after the discovery of very high and rapidly 
variable MeV-GeV fluxes in many OVV (optically violent variable) and 
HP (highly polarized) quasars by the EGRET 
instrument on the Compton Gamma-Ray Observatory (CGRO):  during states of high 
activity, those were measured to exceed the synchrotron fluxes by a 
factor 10 or more \cite{vonmon,fossati}, 
and it was quickly realized that other processes besides SSC can 
be more important 
for $gamma$-ray production \cite{dersch,manbir,SBR,blalev}.

Two families of new models emerged, describing the $\gamma$--ray production:  
these were ERC (external-radiation-Compton) and hadronic models. 
Regarding the external sources of seed soft photons for 
the inverse-Compton process in the ERC models the 
following have been suggested: 
direct disc radiation \cite{dersch}; 
BELR (broad-emission-line-region) \cite{SBR}; 
disc radiation scattered by the gas surrounding the jet \cite{blalev};
and jet synchrotron radiation scattered/reprocessed back to the jet 
by the external gas \cite{ghimad}.
In hadronic models, the $\gamma$-rays are produced 
via synchrotron mechanism by secondary electrons which are much more 
energetic than the electrons accelerated directly.  
They are injected following hadronic processes and  
accompanying pair cascades. The primary reaction in hadronic models
involves collision of ultra-relativistic protons with soft photons
 \cite{manbir} or with the ambient cold protons/ions
\cite{bednarek}. The hadronic scenarios, as alternative for 
the SSC model, have been proposed also to explain TeV radiation, detected 
by atmospheric Cherenkov detectors in a number of low luminosity BL Lac 
objects \cite{laor,rachmes,aha,mucpro}. 

In our review we discuss: 
environment of sub-parsec jets in AGNs; dissipation of jet energy and 
particle acceleration mechanisms; production of non-thermal radiation
and blazar models.

\section*{Sub-parsec jets and their environment}
Typical electromagnetic spectrum of luminous ``non-blazar'' AGNs/quasars 
consists of three distinctive spectral components:  the UV bump and two 
broad (but less luminous) components, the IR and the X-ray one. 
In addition, in the optical-UV range, strong, broad emission lines (BEL) are 
present. UV bumps are produced by central parts of the optically thick
accretion disc; X-rays come from  hot disc coronae, with  possible
additions  from the base of a jet in radio-loud AGNs
\cite{wozniak,hard}; near/mid IR radiation presumably 
comes from dust, located in  geometrically thick molecular torus and 
heated by radiation from the center;
and BELs very likely originate in the disc winds, 
photoionized by UV radiation of an accretion disc 
\cite{emmer,murray,bottorff,nicastro}.
These radiation components form dense radiative
environment, through which jets -- presumably formed in the vicinity of  the 
central black hole -- must pave the way out. Because of Doppler enhancement, 
the relativistic jets interact most effectively with radiation 
incident from the front and from 
side of the jet.  At distances where 
most of blazar non-thermal radiation is  produced 
($ r \sim 10^{17} - 10^{18}$ cm), such radiation is provided  
by the   BEL region  and by the  hot dust. Energy density of BELs on
the  jet axis is
\begin{equation}
 u_{BEL}(r) \simeq f_1 {L_{BEL} \over 4 \pi r_{BEL}^2 c}
\end{equation} 
where $r_{BEL} \sim 0.3 L_{UV,46} $ pc \cite{pet} and
$f_1$ is the correction function which depends on geometry of the BEL region.
If the BEL region forms a narrow torus lying in the disc plane, then
\begin{equation}
f_1  \simeq {1 \over 1 + (r/r_{BEL})^2} 
\end{equation}
Energy density of dust radiation is
\begin{equation}
u_{IR} \simeq \xi_{IR} {4 \sigma_{SB} \over c} T_{dust}^4  
\end{equation}
where 
\begin{equation}
r_{IR} \simeq \sqrt{ {L_d \over 4 \pi \sigma_{SB} T_{dust}^4}}
\end{equation}
and $\xi_{IR}$ is the fraction of the central engine luminosity $L_d$
reprocessed  by hot dust.  

If a luminous AGN possess a powerful jet, it is a radio-loud 
quasar.  Observationally, it is classified as OVV/HP  quasar
if viewed along the jet; as a radio-lobe dominated quasar if viewed at
intermediate angles to the jet axis; and as a radio-galaxy with FR II type
of large scale radio-morphology (edge brightened lobes with hotspots) if
viewed at large angles to the jet axis \cite{ghispad,urpad}. Much less is known
about the sub-parsec environment in very low luminosity AGNs. 
In particular, the accretion process is less-well understood;  we do not know 
about the radiation efficiency of such objects, and to what extent 
optically thick plasma participates in the radiation processes. 
The low luminosity radio-loud AGNs are often accompanied by
strong jet activity.  The objects 
with jets pointing away from the line of sight are 
most likely FR I type radio galaxies with radio morphology characterized by
edge darkened extended radio structures.  FR I radio galaxies probably form 
the parent population for  BL Lac objects \cite{ghispad,urpad}, which, 
in similarity to the OVV/HP quasars, are viewed at small angles 
to the jet axis.  
BL Lac objects, together with the OVV/HP quasars, form 
the {\it blazar} category 
of AGNs. It should be noted, however, that the existence 
of two distinct radio morphologies, FR I and FR II (and 
related bimodal classification of all radio loud 
objects) doesn't necessary imply bimodal distribution of AGN properties.  
Specifically, that  there is a large luminosity overlap between
FR II and FR I objects as well as between BL Lac objects and OVV/HP;
that there is no evidence for  bimodal distribution of equivalent widths 
of emission lines in blazars; that there are some radio galaxies 
which on one side have FR I type radio structure  and on another side 
FR II type radio structure \cite{gopwii};  
and, finally, the global spectral properties 
of blazars seem to form a continuous pattern parameterized by the total blazar 
luminosity \cite{fossati}.

\section*{Jet energy dissipation and particle acceleration}
Rapid variability and the extent of blazar spectra to GeV and TeV energies 
imply particle acceleration {\it in situ}.
The ultimate source of particle energy is the kinetic and intrinsic energy 
of a jet. The simplest and most popular 
model of production of high amplitude short term flares  involves collisions 
of inhomogeneities moving with different velocities down the jet
\cite{SBR,blaz,ghis}. 
If inhomogeneities are ejected by the central engine with bulk Lorentz factors
$\Gamma_f >  \Gamma_s \gg 1$, separated by a distance 
$\Delta r_{ej}$, they start to collide at a distance
\begin{equation}
 r_0 \simeq {c \Delta r_{ej}  \over v_f - v_s} \sim 
{2 \Gamma_f^2 \Gamma_s^2 \over \Gamma_f^2 - \Gamma_s^2} \Delta r_{ej}  
\end{equation}
The collisions are
followed by a formation of forward-reverse  shock structures, with the 
shocked plasma enclosed between the shock fronts and moving with
the bulk Lorentz factor of the contact discontinuity surface, which for
equal rest densities of inhomogeneities is  
$\Gamma \sim \sqrt {\Gamma_f \Gamma_s}$ (for more general cases see 
\cite{daig,lazz}).
As the collision proceeds, 
relativistic particles are accelerated and produce non-thermal radiation.  
For symmetrical inhomogeneities, collisions last
\begin{equation}
t_{coll} = \Gamma t_{coll}' \simeq {\Gamma \lambda' \over v_f'}
\sim { \Gamma \lambda' \over c} \, 
{\Gamma_f^2 + \Gamma^2 \over \Gamma_f^2 - \Gamma^2}
\end{equation}
where $\lambda'$ is the radial width  of the colliding inhomogeneities and
$v_f'$ is the velocity of the faster moving inhomogeneity, both 
as measured in the contact surface frame. After that time the shocks and 
the injection of relativistic particles terminate. Thus, in case of a 
point source, the observer located at $\theta_{obs} \sim 1/\Gamma$ will see
the flare which lasts
\begin{equation}
t_{fl} = t_{coll} (1 - \beta  \cos \theta_{obs}) \simeq 
{t_{coll} \over \Gamma^2} 
\end{equation}
provided electrons cool faster than the collision lasts.
For the flow which is modulated such that $\Delta r_{ej} \sim \lambda'/\Gamma$,
and $\Gamma_f/\Gamma_s \sim $few, Eqs.(5-6) give 
\begin{equation}
t_{coll} \sim {r_0 \over c}
\end{equation} 
i.e. during the collision the source of radiation 
doubles its distance from the central object.  
For sources with finite size, light-travel effects must be included, 
and then the flare time is  
$ t_{fl} \sim ( t_{coll}' + \lambda_{sh}'/c   + a/c )/\Gamma$
where $a$ is the cross-sectional radius of the source and
$\lambda_{sh}'$ is its radial width (the width of
the layer of the shocked gas enclosed between the forward and reverse shock
 fronts). However, because $\lambda_{sh}' \sim (2 v_d'/c)t_{coll}'$,
where $v_d'$ is the shock down-stream velocity,
and because  effectively in the conical jets $a \le r/\Gamma$ 
(no larger area than this can contribute to the observed radiation
due to Doppler beaming), the time scale for a 
source with a finite dimension is of the same order as for the point source.

Efficiency of energy dissipation via collisions of symmetric  inhomogeneities 
is \cite{daig,lazz}
\begin{equation}
 \eta_{diss} \simeq 
1 - {2 \sqrt{\Gamma_f/\Gamma_s} \over 1 + \Gamma_f/\Gamma_s}  
\end{equation}
Some portion  of dissipated energy is converted to the thermal energy of 
the  shocked plasma, and the remaining part is used to accelerate protons 
and electrons up to relativistic energies. Protons are quite efficiently 
accelerated by Fermi  mechanism in shocks 
(see reviews by \cite{drury,blaeic,kui}). 
Time scale of the  acceleration is
\begin{equation}
 t_{acc}' \simeq \zeta \left ( c \over v_{sh}' \right )^2 t_B' 
\end{equation}
where $\zeta$ is the ratio of the mean free path for Fermi scatterings to
the gyro-radius,
$v_{sh}'$ is the velocity of the upstream flow, and 
\begin{equation}
 t_B' = {2 \pi \gamma_p m_p c \over e B'}
\end{equation}
is the gyro-time. The value of $\zeta$ is equal to
$u_B' / k w(k) $, where $w(k)$ is the magnetic energy density per unit
wavenumber $k$ in the turbulent magnetic field,
and $u_B'$ is the total magnetic energy density.
For Kolmogorov turbulence spectrum, $w(k) \propto k^{-5/3}$, i.e.
the smallest
values of $\zeta$ are predicted for most energetic protons \cite{biestr}. 
Protons accelerated  by the Fermi mechanism in strong but
non-relativistic shocks  have a power-law energy distribution with   
a slope corresponding to the equal energy per decade of energy.  

Electrons can be accelerated by shocks as well, but first they must be 
preheated to energies   
$\gamma_{th} \sim (v_u'/c) m_p/m_e$, in order to be scattered freely over the 
shock front and energized between diverging flows, 
where $v_u'$ is the upstream flow velocity in the shock front frame.  
There are several processes which have been suggested to preheat electrons
\cite{hoshino,levinson,romlov,black,mcclements,dieckmann,shimada}. 
The most promising scenario is, where electrons are heated by waves 
induced in the upstream flow by protons  
reflected from the shock front. 

In general, the  slope of energy distribution of preheated electrons can be 
different than the slope of accelerated protons, but those 
accelerated further up to $\gamma \gg \gamma_{th}$ via 
Fermi scatterings will have similar  slope to that of the 
protons.  The large difference 
is predicted, however, for maximum energies, which for electrons
is much stronger limited by radiative energy losses than for protons. 

\section*{Radiation processes}
\noindent
\underline{\it Electron energy losses}

Radiative cooling of relativistic electrons is dominated by synchrotron
mechanism and inverse-Compton process. The rate of radiative cooling
in the source comoving frame, in which electrons are assumed to have
isotropic momentum distribution, is
\begin{equation}
\left \vert {d\gamma \over dt'} \right \vert_{rad} \simeq c_1 \gamma^2 u'
\end{equation}
where $ u' \simeq u_B' + u_{rad<}'$,
$u_B'= {B'}^2/8\pi$ is the magnetic energy density,
$u_{rad<}' = u_{rad}'[\nu < m_e c^2/ h \gamma]$ is the radiation energy 
density 
within the frequency range for which scatterings off electrons with energy 
$\gamma$ are within
the Thomson limit, and  $c_1 = 4\sigma_T/3 m_e c$. 
Comparing time scale of electron radiative cooling,
$t_{e,cool}' \simeq \gamma /\vert d\gamma / dt' \vert_{rad}$, with the
collision time scale, $t_{coll}' \sim \Gamma t_{fl}$, one can find that
energy, above which energy distribution of electrons steepens due to
radiative losses (by unity with respect to the injection slope) is
\begin{equation}
 \gamma_c \simeq {1 \over c_1 \Gamma u' t_{fl}} 
\simeq {24 \over u'} {1 \over (\Gamma/15) (t_{fl}/1 {\rm day})} 
\end{equation}

Radiation energy density is contributed by both local emission
and by external radiation fields. The dominant local contribution to
$u_{rad<}'$ for $\gamma \gg 1$ is provided by synchrotron radiation: 
\begin{equation}
 u_{syn<}' \simeq 
{ L_{syn<}  \over 2 \pi a^2 c \Gamma^4}
\end{equation}
where $L_{syn<}$ is
the apparent luminosity of synchrotron radiation within the 
Thomson limit ($\nu < \Gamma  m_e c^2/ h \gamma$).
Contribution from the external radiation fields to $u_{rad}'$ is 
\begin{equation}
 u_{ext}' = {1 \over c} \int I' d \Omega' =
{1 \over c} \int I {\cal D}_{in}^{-2} d \Omega 
\end{equation}
where $I$ is the intensity of the incoming external radiation, 
${\cal D}_{in} = 1 / \Gamma (1 - \beta \cos \theta_{in})$, and 
$\theta_{in}$ is the angle between the jet axis and an incoming ray.
 Provided that the dependence
of  $I$ on $\theta_{in}$ is much weaker than the dependence of ${\cal D}_{in}$
on $\theta_{in}$, the $u_{ext}'$ is dominated by those external radiation 
fields which at a distance of the flare production contribute significantly
from the side and the front of the jet \cite{siksol}.  
Such radiation fields are provided by BEL region and by infrared radiation by 
hot dust, and from  Eq. (15) 
\begin{equation}
 u_{ext}'(r) \simeq f_2 \Gamma^2 u_{ext}(r)  
\end{equation}
where  for the external sources located in the narrow torus at $r_{ext}$
\begin{equation}
f_2 \simeq  \left (1 - {r/r_{ext} \over \sqrt{1 + (r/r_{ext})^2}} \right )^2
\end{equation}
Noting that distance of the flare production,
\begin{equation}
r_{fl} \sim 
c t_{fl} \Gamma^2 \sim 6 \times 10^{17} (t_{fl}/1 {\rm day}) (\Gamma/15)^2 
\, {\rm cm}
\end{equation}
is on the order of $r_{BEL}$, the value of $f_2$ can be sometimes 
significantly lower than unity. For scatterings of IR radiation,
because $r_{dust} \gg r_{fl}$, we have $f_2 \simeq 1$.
Noting that $\nu_{BEL} \simeq 10^{15}$ Hz, 
$\nu_{IR} \sim 3 k T_{dust}/h \simeq 6 \times 10^{13} T_3$ Hz, 
and that $\nu_{ext}' \simeq {\cal D}_{in}^{-1} \nu_{ext} = 
\sqrt{f_2} \Gamma \nu_{ext}$, 
one can find that BELs are  scattered  within the Thomson limit by 
electrons with energies 
\begin{equation}
\gamma < {m_e c^2 \over \sqrt {f_2} \Gamma  h \nu_{BEL}} 
\sim {7.9 \times 10^3 \over \sqrt{f_2} (\Gamma/15)}
\end{equation}
while the dust radiation is scattered within the Thomson limit
by electrons with energies
\begin{equation}
\gamma < {m_e c^2 \over \sqrt \Gamma  h \nu_{IR}} 
\sim {1.3 \times 10^5 \over T_3 (\Gamma/15)} \, .
\end{equation}

\noindent
\underline{\it Proton energy losses}

Protons lose energy  via synchrotron and inverse Compton radiation
at a rate
\begin{equation}
\left \vert {d\gamma_p \over dt'} \right \vert \simeq c_2 \gamma^2 u'
\end{equation}
where $c_2 = (m_e/m_p)^3 c_1$.  
 
Comparing the time scale of proton radiative cooling,
$t_{p,cool}' = \gamma_p /\vert d\gamma_p/ dt' \vert$, with the flare 
time scale, one can find that only protons with energies larger than
\begin{equation}
\gamma_{p,c}^{(rad)} \simeq {1 \over c_2 \Gamma u' t_{fl}} 
\simeq {1.5 \times 10^{11} \over u'} 
{1 \over (\Gamma/15) (t_{fl}/1 {\rm day})} 
\end{equation}
can radiate efficiently.  

Protons can also lose energy  via inelastic collisions with photons and cold 
ions. Collisions with photons can lead
to meson production (photo-meson process) and to direct e$^{+}$e$^{-}$ pair
production. Time scales of proton energy losses in these processes are
\cite{BRS}: 
\begin{equation}
t_{p\gamma}^{(\pi)\prime} 
\simeq {5.6 \times 10^{17} \over n_{>}^{(\pi)\prime}}  
\end{equation}
\begin{equation}
t_{p\gamma}^{(e)\prime} \simeq 
{6.7 \times 10^{19} \over n_{>}^{(e)\prime}} 
\end{equation}
\begin{equation}
t_{pp}' \simeq {2.2 \times 10^{15} \over  n_p'} 
\end{equation}
for photo-meson production, pair production, and $pp$ collisions, 
respectively, where
$n_{>}^{(\pi)\prime} = \int_{ \nu_{th}^{(\pi)\prime}} n_{\nu}' \, d \nu'$,
$n_{>}^{(e)\prime} = \int_{ \nu_{th}^{(e)\prime}} n_{\nu}' \, d \nu'$, and the 
threshold photon energies for $p\gamma$ collisions are
$ \nu_{th}^{(\pi)\prime} \simeq 
m_{\pi}c^2/h \gamma_p \simeq 3 \times 10^{22}/\gamma_p$,
and $ \nu_{th}^{(e)\prime} \simeq 2m_{e}c^2/h \gamma_p \simeq 2 \times 10^{20}/\gamma_p$.

\section*{Blazar models}
Depending on the dominant mechanism considered for the 
$\gamma$-ray production, blazar models can be divided into three groups: 

\noindent
- SSC models, where production of $\gamma$-rays is dominated by Comptonization
of locally emitted synchrotron radiation;

\noindent
- ERC models, where production of $\gamma$-rays is dominated by Comptonization 
of external radiation fields; 

\noindent
- hadronic models, where production of $\gamma$-rays is dominated by 
synchrotron radiation of proton-initiated-cascades (PIC), and/or by
synchrotron radiation of protons and muons.

\noindent
These models are discussed below  as applied to OVV/HP quasars and TeV-BL Lac 
objects.
 
\subsection*{SSC models}
The simplest SSC model is the one-component version, 
where the radiation observed at any given moment
is produced by a source/shock moving along a jet. In the SSC  model, the 
frequency of the $\gamma$-ray luminosity peak is located at
\begin{equation}
\nu_{H} \sim \cases {\gamma_m^2 \nu_{L} & 
if $h\nu_{L} > \gamma_m m_e c^2/\Gamma$ \cr
\gamma_m m_e c^2/h &otherwise \cr} 
\end{equation}
where
\begin{equation}
\nu_{L} \sim c_B \gamma_m^2 \Gamma B'
\end{equation}
is the location of the synchrotron peak,
$\gamma_m$ is the location of the break in 
the electron energy distribution, and $c_B = (2/3\pi)(e/m_e c)$.  
These equations together with the ratio of the luminosity peaks,
\begin{equation} 
{\nu_H L_{\nu_H} \over \nu_L L_{\nu_L}} \simeq {u_{syn,<}' \over u_B'}
\end{equation}
allow to derive $B'$, $\Gamma$, and  $\gamma_m$.
\smallskip

\noindent
\underline{\it SSC in OVV/HP quasars}
\smallskip

The SSC process can dominate $\gamma$-ray production in OVV/HP quasars if
\begin{equation}
 {L_{SSC} \over L_{ERC}} \sim {u_{syn}' \over u_{ext}'} > 1
\end{equation}
This can be satisfied in OVV/HP quasars only if
geometry of the BEL region is such that $f_2 < 10^{-2}$ and if temperature
of dust is lower than $300$ K, or that $\Gamma < 5$.
First condition is difficult to satisfy, because
from reverberation mapping, $r_{BEL} \sim r_{fl}$;  
second condition is in contradiction with observations of near/mid IR 
radiation in AGNs of radio-lobe dominated quasars. Finally, $\Gamma < 5$
contradicts with VLBI data. Furthermore, one-component SSC model predicts 
X-ray spectra which are too soft.  

\smallskip

\noindent
\underline{\it SSC in TeV-BL Lac objects}
\smallskip

Much more promising are SSC models as applied to BL Lac objects, in
particular to the BL Lac objects which are TeV sources 
\cite{taka,tav98,KRM,bedpro}. In these objects, 
Klein-Nishina effects are important. Using Eqs. (26-28), 
one can find that typically $B' \sim 0.3$ Gauss and 
$\gamma_m \sim 3 \times 10^4$
\cite{tav98}. Using these numbers, Eq. (13) gives $\gamma_c \sim \gamma_m$. 
This suggests that spectral breaks at peaks are
due to effects of cooling;  furthermore, the fact that $\gamma_m$ 
 is only by a factor few lower than 
$\gamma_{max} \sim h \nu_{H} /\Gamma m_e c^2$ explains why 
spectral components in TeV blazars are hard almost up to highest
frequencies. Finally,
noting that maximum electron energies presumably are determined by the 
balance of the acceleration time scale and radiative cooling time scale, 
having $t_{coll}' \sim t_{cool}'$ provides natural explanation
for existence of ``histeretic''  variability patterns  observed with 
both signs 
in the X-ray band in the spectral index vs. flux diagram 
\cite{taka,taka2000,KRM,foss}.

\smallskip

\subsection*{ERC models}
\smallskip

In the ERC one-component model the high energy peak is located at
\begin{equation} 
\nu_{H} \sim \xi_1 \Gamma^2 \gamma_m^2 \nu_{ext} 
\end{equation}
while ratio of the peak luminosities is 
\begin{equation}
{\nu_H L_{\nu_H} \over \nu_L L_{\nu_L}} \simeq {u_{ext}' \over u_B'}
\end{equation}

\noindent
\underline{\it ERC in OVV / HP blazars}
\smallskip

When applying ERC model to Q-blazars, it is better not to rely on
relation (27). This is not only because
$\nu_{L}$ is located in these objects in the spectral range which 
observationally is poorly covered (far-IR band), but also because there are 
some observational indications that far-infrared radiation can be strongly 
contaminated by radiation produced at larger distance in a jet than the 
high energy flares. These indications are: the low energy break of the
synchrotron component is at lower frequencies than
the break due to synchrotron-self-absorption on sub-parsec scales \cite{SBR};
and there is  clear trend of decreasing amplitude of the flare with decreasing
frequency in the synchrotron component \cite{edel,edelmal,weh}.
However, in order 
to close a system of equations describing the model, one has to replace 
one equation by another.  Such relationship can be derived by 
assuming that $\gamma_m = \gamma_c$ \cite{blaz}.  
Another useful approximation is to treat
as a known observable $\Gamma$, instead of $u_{ext}$. This is because due to
unknown geometry of the BEL region and unknown maximum temperature of dust,
model errors which can result from  using uncertain values of 
$f_2$ and $T_{dust}$ can be
larger, than resulting from uncertain value of $\Gamma$. Alternatively,
instead of treating $\Gamma$ as known, one can make an assumption 
about equipartition 
of magnetic fields with relativistic particles. Our preliminary results show 
that for $\Gamma \sim 15$ both approaches give similar results.  
Using observables typical of OVV/HP blazars, namely:  
$\nu_{H} = 10^{21}$ Hz; $\nu_H L_{\nu_H}/\nu_L L_{\nu_L} = 10$;
$t_{fl} = 1$ day; and $\Gamma=15$, and assuming that the high energy luminosity
is dominated by ERC(BEL), we obtain 
$B' \sim 1$ Gauss; $\gamma_c \sim 70$; $\nu_L \sim 2 \times 10^{11}$ Hz 
(which actually is below the synchrotron-self-absorption frequency); 
$\gamma_{max} \sim 4 \times 10^3 \sqrt{\nu_{syn,max}/10^{15}{\rm Hz}}$;
and $f_2 \sim 0.2$. 
 
Model  parameters derived in this manner can be used as ``first 
approximation'' parameters for dynamical models intended for 
a study of time evolution of electron energy distribution.  
Such models have been recently developed using two kinds of
approximations: one is a 
spherical source approximation by \cite{chiagis,ghis} and another is 
a thin shell approximation by \cite{blaz}.  
Results of both demonstrate the ability of ERC models to explain the observed 
spectra and flares of OVV/HP quasars. In addition, these models
support the earlier
predictions that the SSC radiation component, despite its lower luminosity than
the ERC component, can dominate the soft/mid X-ray bands \cite{inoue,kubo}.

\smallskip
\noindent
\underline{\it ERC in TeV-blazars}
\smallskip

Very little is known about radiative environment in these objects
and, so, energy density of external radiation fields must be treated
as a free parameter. But there are 
interesting constraints regarding  the level of the IR  radiation from dust.
This radiation must be very weak, otherwise  TeV $\gamma$--rays 
would be absorbed 
in the pair production process \cite{CFR}. However, production of 
TeV component by Comptonization of radiation of dust is still feasible,
provided $\Gamma >15$.

\subsection*{Hadronic models}

Hadronic models, which are motivated by theories of particle 
acceleration, suffer difficulties in explaining the observed electromagnetic
spectra and/or  short term variability.  
Proposed for the OVV/HP quasars, models involving 
proton induced cascades (PIC) \cite{manbir} 
predict X-ray spectra which are softer than observed.
  They also require 
fine-tuning of model parameters in order to have the location of the 
radiation 
deficiency (the "dip" between the low energy and high energy peaks in the 
broad-band spectra) 
located at the right frequency in all OVV/HP blazars. Furthermore, 
there are already several examples
of OVV/HP blazars, where X-ray spectra have slopes $\alpha_X < 0.5$
\cite{fabian,reeves,tav00}, which
cannot be explained in terms of the synchrotron radiation mechanism.
This is because synchrotron radiation in the X-ray band  requires such 
energetic  electrons, that they must cool efficiently and, therefore, produce
spectra with $\alpha_X > 0.5$, even if injected  mono-energetically.

From unknown us reasons, no contribution of pp$\to$ pe$^{+}$e$^{-}$ was
taken into account in PIC models. The process is about 100 times less efficient
per photon than photo-meson one, but this is compensated by 
a much larger number of photons above the threshold energy, which for pair 
production is $\sim 100$ lower than for photo-meson process.

Another hadronic model proposed for OVV/HP blazars is based on the process of
inelastic collisions of 
ultrarelativistic protons with the cold ions. Such process in order
to be efficient
requires very dense targets and huge bulk Lorentz factor ($\Gamma \gg 10$) 
of streaming protons. 
In the paper \cite{bednarek} it is 
proposed that the dense target is provided by the walls of the funnel formed
by geometrically thick disc. The main problem with this model we can envisage
is overproduction of soft X-rays, which would be produced due to
Comptonization of accretion disk UV  photons by cold electrons in a jet
This effect can be avoided if in the vicinity of the black hole,
up to at least 100 gravitational radii, jet is not yet collimated
\cite{SMMP}. But then beamed blazar radiation must be produced at larger
distances, probably at $r \sim 10^{17-18}$ cm as the flare time scales 
suggest.

Hadronic models have been proposed also for TeV-blazars.
In these objects the  soft spectra of PIC models contrast with
the very hard observed spectra even more. This problem is  avoided
in proposed recently the proton-synchrotron model \cite{aha,mucpro}. In such 
a model high energy spectra are produced directly by protons, just via
synchrotron radiation. The model predicts hard spectra and is also attractive,
because can explain constant slope of high energy tail of TeV component
during high amplitude flux changes \cite{aha}.
In the model, maximum proton energies 
$\gamma_{p,max} > 10^{10}/\sqrt{B'/100 {\rm G}}$ are
required in order to satisfy condition 
$t_{p,syn}'(\gamma_{p,max}) \le t_{fl} \Gamma$. 

Another version of the hadronic model has been suggested by Rachen \&
M\'esz\'aros \cite{rachmes}. In their model the bulk Lorentz factor is 
assumed to be rather low ($\Gamma \sim 3$) and  protons with 
$\gamma_{p,max} > 10^{10}$ lose comparable energies
via photo-meson process as via synchrotron mechanism. Following the former,
the muons are produced with such energies, that before decay they undergo
significant synchrotron energy losses, dominating production of TeV peak.
The weak point of both above hadronic models is that in order to accelerate 
protons up to $\gamma_{p,max} > 10^{10}$ energies, 
the quantities $\zeta$ and $v_{sh}'/c$ (see Eq. [10]) must be pushed to 
extremes, i.e., should have values 1.

\section*{Summary}
As analysis of jet energetics and X-ray observations of blazars indicate,
plasma  in sub-parsec jets can contain from few  up to tens of e$^{+}$e$^{-}$ 
pairs per proton \cite{sikmad,blaz}. With such a  number of pairs 
energy flux of sub-parsec jets is still dominated by protons, 
and, therefore, the structure of shocks is determined by proton plasma.
Then, one should expect very efficient shock acceleration of
protons, with  maximum energies as high as $\gamma_p \sim 10^9-10^{10}$,
as limited by size of the source or balanced by energy losses.
Furthermore, very large values 
of $\zeta$ (ratio of the  mean free path for Fermi scatterings to gyro-radius)
for electrons, $10^{3-6}$, as derived for maximum electron energies 
assuming that for them $t_{cool}' = t_{acc}'$,  
seem to be consistent with the presence of ultrarelativistic protons,
provided the turbulent magnetic field has a Kolmogorov spectrum \cite{biestr}.
On the other hand, very hard X-ray spectra and  short 
variability time scales put severe constraints on 
radiative role of protons. Does this  contradict with theoretical predictions 
about very efficient acceleration of protons? Not necessary. 

The average proton
cannot reach more random energy in the shock than 
$\bar E_p \sim (\Gamma_u'-1)m_pc^2$, where
$\Gamma_u'$ is the bulk Lorentz factor of the upstream flow as measured in the 
rest frame of the shock front. Intrinsic shocks are at most mildly
relativistic, and therefore $\bar E_p < m_p c^2$ is expected.
Comparing this with the average energy of electrons which, as deduced from hard
X-ray spectra of some OVV/HP quasars, is in the range $10-100$ MeV, and noting
that limited energetics of jets combined with electron emissivity implies 
$n_e/n_p$ to be at least of the order of 10, one can find that
fraction of  energy dissipated in the shock and used to accelerate
electrons does not have to be much lower than energy channeled to protons.
This, combined with the fact that radiative efficiency of electrons
is much larger than of protons, can explain negligible contribution
of the proton related processes to the observed spectra.    

\smallskip
\smallskip
{\bf Acknowledgments:} This work has been supported in part by the Polish
KBN grant 2P03D 00415.

\end{document}